%% file: p.tex
\documentclass[letterpaper,twocolumn,10pt]{article}
\usepackage{usenix}

\input{pkgs}

\newcommand{\sys}{\mbox{\textsc{TrustDesc}}\xspace}
\newcommand{\slicemin}{\cc{SliceMin}\xspace}
\newcommand{\dynver}{\cc{DynVer}\xspace}
\newcommand{\descgen}{\cc{DescGen}\xspace}

\definecolor{diffred}{RGB}{160, 0, 0}
\definecolor{diffgreen}{RGB}{0, 130, 0}

\input{cmds}
\input{rev}

\begin{document}

\input{hdr}
\date{}

\maketitle

\input{abstract}

\input{intro}

\input{background}
\input{problem}
\input{design}

\input{eval}
\input{discuss}
\input{conclusion}

\balance
\bibliographystyle{plain}
\bibliography{p,conf,refs-pi}

\end{document}

%% file: pkgs.tex
\usepackage{xcolor}
\usepackage{color,colortbl}
\usepackage{url}

\usepackage{hyperref}
\usepackage{amsopn}
\usepackage{amsthm}
\usepackage{amssymb}
\usepackage{microtype}
\usepackage{xspace}
\usepackage{multirow}
\usepackage{makecell}
\usepackage{underscore}
\usepackage{relsize}
\usepackage[T1]{fontenc}
\usepackage{enumitem}
\usepackage{subcaption}
\usepackage[labelfont=bf,font={small,bf},skip=5pt]{caption}
\usepackage[linesnumbered,lined,ruled,vlined]{algorithm2e}
\usepackage{setspace}
\usepackage{balance}
\usepackage{pifont}

\usepackage{booktabs}
\usepackage{fp}
\usepackage{siunitx}

\usepackage{listings}

\makeatletter
\@namedef{ver@lineno.sty}{9999/12/31}
\@namedef{opt@lineno.sty}{}
\makeatother
\usepackage{minted}

%% file: cmds.tex
\newcommand{\cc}[1]{\mbox{\smaller[0.5]\texttt{#1}}}

\fvset{fontsize=\scriptsize,xleftmargin=9pt,numbers=left,numbersep=4pt}

\input{code/fmt}
\newcommand{\figrule}{\hrule width \hsize height .33pt}
\newcommand{\coderule}{\vspace{-0.8em}\figrule\vspace{.2em}}

\def\Snospace~{\S{}}

\definecolor{darkgreen}{rgb}{0, 0.5, 0}
\definecolor{mygray}{gray}{0.9}

\newif\ifdraft\drafttrue
\newif\ifnotes\notestrue
\ifdraft\else\notesfalse\fi

\input{glyphtounicode}
\newcolumntype{R}[1]{>{\raggedleft\let\newline\\\arraybackslash\hspace{0pt}}p{#1}}

\newcommand{\squishlist}{
\begin{itemize}[noitemsep,nolistsep]
  \setlength{\itemsep}{-0pt}
}
\newcommand{\squishend}{
  \end{itemize}
}

\usepackage{tikz}
\newcommand*\WC[1]{%
\begin{tikzpicture}[baseline=(C.base)]
\node[draw,circle,inner sep=0.2pt](C) {#1};
\end{tikzpicture}}

\newcommand{\PP}[1]{%
\vspace{2px}%
\noindent{\bf #1.}%
}

\newcommand{\boxbeg}{
\vspace{2px}
\noindent\begin{tabular}{|l|}\hline
  \rowcolor{gray!20}
\begin{minipage}{3.15in}
\vspace{3px}
\noindent
}

\newcommand{\boxend}{
\vspace{3px}
\end{minipage}\\ \hline
\end{tabular}
}

\newcommand{\etal}{\textit{et al}.\xspace}
\newcommand{\ie}{\textit{i}.\textit{e}.\xspace}
\newcommand{\eg}{\textit{e}.\textit{g}.\xspace}

\newcommand{\ExternalLinkX}{%
    \tikz[x=1.2ex, y=1.2ex, baseline=-0.05ex]{%
        \begin{scope}[x=1ex, y=1ex]
            \clip (-0.1,-0.1)
                --++ (-0, 1.2)
                --++ (0.6, 0)
                --++ (0, -0.6)
                --++ (0.6, 0)
                --++ (0, -1);
            \path[draw,
                line width = 0.5,
                rounded corners=0.5]
                (0,0) rectangle (1,1);
        \end{scope}
        \path[draw, line width = 0.5] (0.5, 0.5)
            -- (1, 1);
        \path[draw, line width = 0.5] (0.6, 1)
            -- (1, 1) -- (1, 0.6);
        }
    }
\newcommand{\ExternalLink}[1]{\href{#1}{\ExternalLinkX}}

\theoremstyle{definition}

\SetVlineSkip{.3mm}

\SetCommentSty{mycommfont}

\usepackage{harveyballs}

\makeatletter
\newcommand{\removelatexerror}{\let\@latex@error\@gobble}
\makeatother

\makeatletter
\patchcmd{\@algocf@start}
  {-1.5em}
  {0pt}
  {}{}
\makeatother

\SetAlCapNameFnt{\small}
\SetAlCapFnt{\small}

\newcommand{\bmidrule}{\arrayrulecolor{black}\midrule}

\newcommand{\second}{$^{\prime\prime}$\xspace}

\usepackage{siunitx}
\usepackage{diagbox}
\usepackage{slashbox}

%% file: rev.tex
\gdef\therev{c13490e}
\gdef\thedate{2026-01-21 10:21:54 -0500}

%% file: hdr.tex
\title{\sys: Generating Trusted Tool Descriptions for \\ LLM-integrated Applications}
\title{\sys: Preventing Tool Poisoning in LLM Applications \\via Trusted Description Generation}

\ifdefined\DRAFT
 \pagestyle{fancyplain}
 \lhead{Rev.~\therev}
 \rhead{\thedate}
 \cfoot{\thepage\ of \pageref{LastPage}}
\fi


\author{
 \rm 
 Hengkai Ye \ \ \ \ \ \ 
 Zhechang Zhang \ \ \ \ \ \ 
 Jinyuan Jia \ \ \ \ \ \ \
 Hong Hu
\\\\
 \textit{The Pennsylvania State University}
}

%% file: abstract.tex
\begin{abstract}

\noindent
Large language models (LLMs) 
increasingly rely on
external tools to perform time-sensitive tasks
and real-world actions.
While tool integration
expands LLM capabilities,
it also introduces a new
prompt-injection attack surface:
tool poisoning attacks (TPAs).
Attackers manipulate tool descriptions
by embedding malicious instructions
(explicit TPAs)
or misleading claims
(implicit TPAs)
to influence model behavior and tool selection.
Existing defenses mainly
detect anomalous instructions and
remain ineffective against implicit TPAs.

In this paper,
we present \sys, 
the first framework
for preventing tool poisoning
by automatically generating
trusted tool descriptions
from implementations.
\sys derives implementation-faithful descriptions
through a three-stage pipeline.
\slicemin performs
reachability-aware static analysis
and LLM-guided debloating
to extract minimal tool-relevant code slices.
\descgen synthesizes descriptions
from these slices
while mitigating misleading or
adversarial code artifacts.
\dynver refines descriptions
through dynamic verification
by executing synthesized tasks
and validating behavioral claims.
We evaluate \sys
on 52 real-world tools
across multiple tool ecosystems.
Results show that 
\sys produces accurate tool descriptions
that improve task completion rates 
while mitigating implicit TPAs at their root,
with minimal time and monetary overhead.

\end{abstract}

%% file: intro.tex
\section{Introduction}
\label{s:intro}

Large language models (LLMs) have 
been widely integrated into
real-world applications
due to their remarkable performance
in natural language understanding, 
reasoning and generation~\cite{llm-general-1,
llm-general-2, llm-general-3}.
Despite these advances,
standalone LLMs still
face limitations in
handling time-sensitive tasks
and performing concrete actions.
To address these limitations,
recent work
augments LLMs with external tools%
~\cite{das2024mathsensei, yuan2023craft,qu2025tool},
allowing models to
retrieve up-to-date information,
execute commands,
and interact with external systems.
Consequently,
tool integration
has become a fundamental component
of modern LLM-based applications.
For example,
Gemini~\cite{team2023gemini}
integrates with a suite of Google services,
such as YouTube and Gmail,
enabling the model
to support diverse user tasks%
~\cite{gemini-app}.
%
%

An LLM tool usually
contains two components:
the executable code
and a semantic description.
The executable code performs the 
actual operations, like retrieving data.
It resides on the application side,
remaining invisible to the LLM.
In contrast,
the description specifies
the tool's functionality
and input schema.
By embedding tool descriptions
into the model context,
LLMs identify available capabilities 
and select proper tools
to assist with user tasks.
%
%
Therefore,
tool descriptions
form a critical trust boundary
that influences decision-making
and execution behavior
in LLM-integrated applications.

Current tool design assumes that
tool descriptions are benign and
faithful to their implementations.
However,
violations of this assumption
introduce a new attack surface,
known as \emph{tool poisoning attacks (TPAs)}%
~\cite{greshake2023not,ignore_previous_prompt,liu2024formalizing,liu2023prompt,tool-poisoning-1,tool-poisoning-2,shi2025prompt,li2025dissonances}.
In \emph{explicit} TPAs,
attackers embed malicious instructions
into tool descriptions.
After being loaded into the model context,
they steer the system 
to execute unauthorized or unintended actions%
~\cite{tool-poisoning-1,tool-poisoning-2}.
In \emph{implicit} TPAs,
attackers craft
benign-looking but exaggerated descriptions
(\eg, ``best'' and ``efficient'') to
bias the LLM's tool-selection process%
~\cite{shi2025prompt, li2025dissonances}.
As a result,
the LLM may prefer
attacker-controlled tools,
systematically altering execution behavior
and potentially producing unsafe outcomes,
even without explicit malicious instructions.


%

Existing defenses against prompt injection%
~\cite{chen2025secalign,wu2025instructional,hung2025attention,abdelnabi2025get,wallace2024instruction,chen2025struq,liu2025datasentinel,liu2024formalizing}
primarily target explicit attacks,
but remain ineffective against implicit ones.
These defenses typically
detect anomalous or malicious instructions
embedded in prompts or tool descriptions%
~\cite{mcp-scanner-1,mcp-scanner-2,
mcp-scanner-3, mcp-scanner-4
}.
For example,
SecAlign~\cite{chen2025secalign}
improve LLM robustness by 
fine-tuning models on datasets
containing labeled prompt-injection attacks.
%
%
%
However,
implicit attacks 
present benign-looking yet misleading descriptions
without any explicit malicious signals,
enabling them to evade automated detectors and
rule-based scanners.
Detecting implicit TPAs
requires reasoning about
whether a tool's description
faithfully reflects its implementation,
a challenging task
even for human experts
without careful code inspection.


We observe that
the core vulnerability
arises not from the tool implementation,
but from discrepancies
between what the description claims
and what the implementation actually performs.
If this inconsistency is eliminated,
tool poisoning attacks
can be mitigated at their root.
Our analysis of
real-world attacks shows
that adversaries rarely embed
malicious code in tools,
likely due to the effectiveness of
modern malware-detection techniques~\cite{brown2024automated,
bensaoud2024survey,kolbitsch2009effective}.
Instead,
attackers predominantly
target the descriptive layer,
which lacks comparable integrity guarantees.
Based on these observations,
we believe that
a tool's source code provides
a trustworthy ground truth
from which faithful tool descriptions
should be derived.



Given their strong capabilities
in understanding and summarizing program code%
~\cite{llm-code-summarization-1,llm-code-summarization-2,deepwiki},
LLMs are natural choices
for generating tool descriptions
from implementations.
However,
applying existing LLM-based
code summarization techniques
faces three non-trivial challenges.
First,
real-world LLM toolsets
often implement multiple tools
within a shared and tightly coupled codebase
(\eg, the \cc{filesystem} toolset contains 14 tools).
Prompting an LLM
with the entire codebase 
incurs significant overhead 
and increases reasoning complexity.
Second,
even when relevant functions are identified,
extracted code may still contain
irrelevant logic
that distracts the LLM
and degrades summary accuracy%
~\cite{shi2023large, wueasily, hwang2025llms}.
Finally,
due to hallucination issues%
~\cite{ji2023survey,huang2025survey}, 
LLMs may generate incorrect claims
when reasoning complex program semantics,
undermining the reliability of 
generated tool descriptions.

To address these challenges,
we design \sys,
the first framework that 
automatically generates
implementation-faithful tool descriptions
from tool implementations.
\sys consists of three components:
\slicemin, \descgen and \dynver.
Given a complete codebase,
\slicemin extracts tool-relevant code slices
through reachability-aware static analysis
and constructs a call graph
for each tool.
It then leverages an LLM
to prune unreachable and irrelevant logic
based on concrete call sites,
iteratively refining the call graph.
Based on the resulting minimal code slices,
\descgen generates initial tool descriptions.
During the generation,
it removes code comments and docstrings,
truncates long function/variable identifiers,
and adopts an LLM-based method
to mitigate misleading or adversarial artifacts 
embedded in source code. 
To address hallucinations
and semantic errors,
\dynver performs dynamic verification
by decomposing descriptions
into verifiable tasks,
executing them,
and recording execution logs.
It uses an LLM-based judge%
~\cite{wang2025mcp,gu2024survey,zheng2023judging}
to analyze these logs
to validate behavioral claims.
After removing statements
that fail dynamic verification,
\sys produces high-quality, trusted tool descriptions
for LLM-integrated applications.

%
%

We implement \sys
using 2,377 lines of Python code
to automatically generate trusted tool descriptions
for model context protocol (MCP) servers.
MCP servers have become a widely adopted
deployment model for real-world LLM tools
and represent natural and impactful targets
for securing tool descriptions.
The current implementation supports
MCP servers developed in Python and TypeScript,
the two most popular programming languages
in MCP ecosystems.
Although our prototype currently
targets Python and TypeScript,
the design of \sys is largely language-agnostic.
We can extend it to other programming languages
by integrating corresponding parsing
and execution frameworks.
Therefore,
\sys is applicable to a broader class
of LLM tool implementations
beyond the MCP ecosystem.

We systematically evaluate 
the accuracy, cost, effectiveness,
and robustness of \sys.
First,
we evaluate 208 tasks
across 52 tools from 12 MCP servers
using both original and \sys-generated descriptions.
Using task success rate (TSR)
as the accuracy metric,
\sys improves task success by 4.3\%
on average,
demonstrating that 
automatically generated descriptions
are highly faithful to
tool implementations.
Second,
we evaluate the cost of
generating and consuming trusted descriptions.
When powered by Gemini-3-Flash, 
\sys requires only 
\$0.013 and 25.7\,s to generate 
a trusted description.
During real-world task execution,
\sys increases monetary cost
by only 4\% and latency by 0.2\%,
on average,
while in some cases
reducing overall cost.
Third,
to evaluate effectiveness
under tool competition,
we introduce low-quality tool variants
that remove security checks,
or disable one to two key functionalities.
\sys reduces the selection rate
of these low-quality tools 
to 41.6\%, 18.8\%, and 14.5\%,
respectively,
demonstrating the improved
tool-quality discrimination.
Finally,
we evaluate \sys's robustness
against adaptive attacks,
which introduce misleading identifiers
to tool implementation
to bias description generation.
Across 15 attack iterations, 
the success rate fluctuates 
between 44.7\% and 67.4\%, 
without a stable upward trend,
indicating that \sys 
remains resilient to
iterative adversarial strategies.

In summary, we make the
following contributions.

\begin{itemize}[topsep=2pt,partopsep=4pt,itemsep=2pt,parsep=0pt,leftmargin=8pt]

\item We propose \sys,
the first framework
that automatically generates
trusted tool descriptions
from implementations,
eliminating description-based attacks
at their root.
    
\item We introduce
a semantics-aware description
generation pipeline
that bridges program analysis and LLM reasoning,
enabling reliable code-to-description translation while mitigating misleading code artifacts and hallucinations.

\item We conduct a comprehensive evaluation
on 52 real-world tools across 12 MCP servers,
showing that \sys improves task
success rates by 4.3\% on average,
reduces the selection of low-quality tools,
and generates trusted descriptions with
minimal cost and runtime overhead.
    
\end{itemize}

%% file: background.tex
\begin{figure}[t]
    \centering
    \includegraphics[width=\columnwidth]{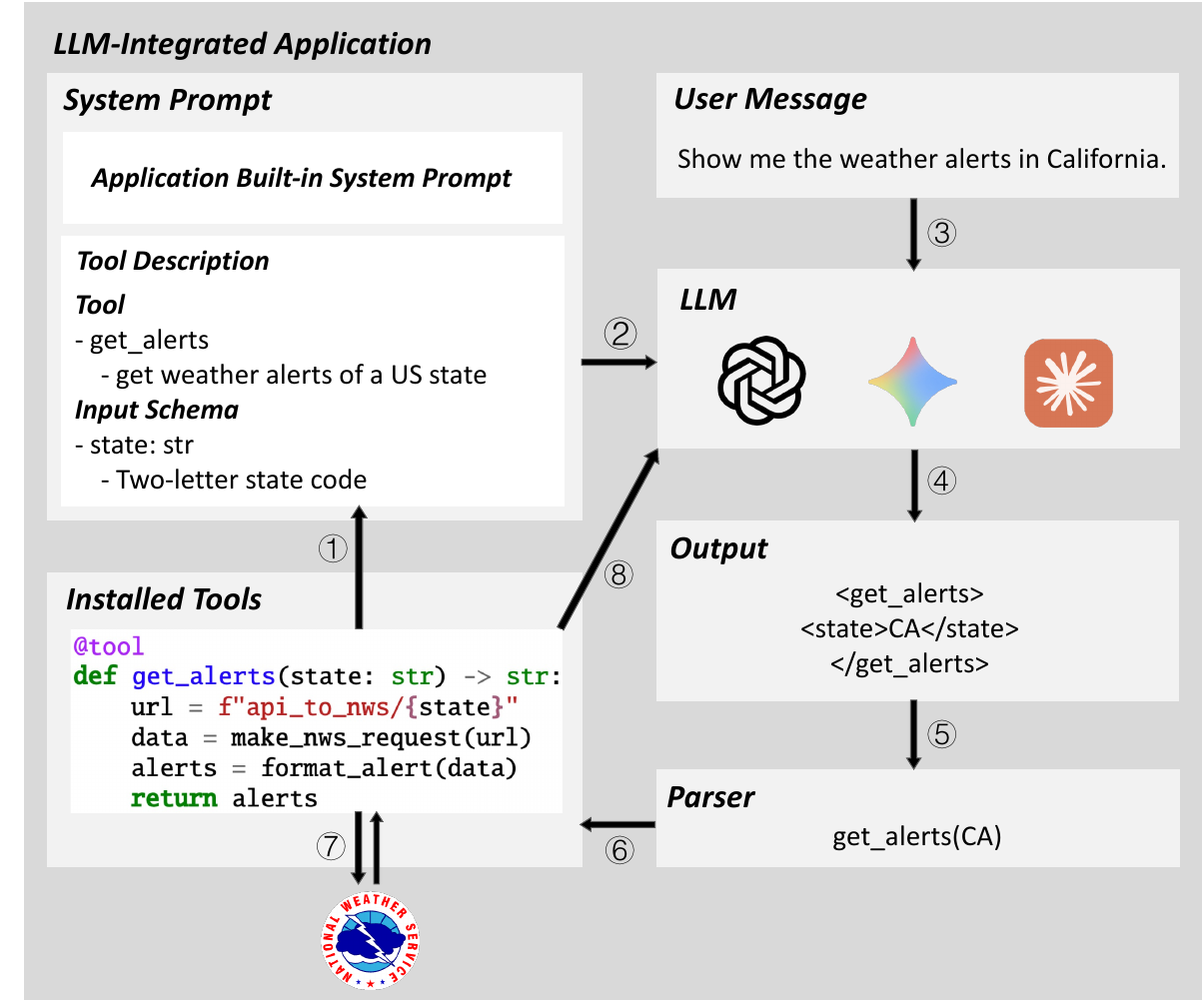}
    \caption{LLM tool-use workflow.
    \textnormal{The tool description plays a critical role
    for LLM-based applications to select the tool.}}
    \label{f:tool_use}
\end{figure}

\section{Background}
\label{s:bg}

In this section,
we provide the background necessary
to understand LLM-based tool use
and tool poisoning attacks.

\subsection{LLM Tool and MCP}
\label{ss:tool-use}

LLMs have demonstrated impressive capabilities
in natural language understanding and generation%
~\cite{llm-general-1,llm-general-2,llm-general-3}.
Augmenting LLMs with external tools
further extends these capabilities
by enabling access to
up-to-date information and
execution of concrete actions.
Tool integration has become 
a core component
of modern LLM applications.
Mainstream systems,
such as ChatGPT, Gemini, and Claude,
integrate web-search tools
to retrieve time-sensitive information,
while coding agents
(\eg, Cline~\cite{cline})
provide built-in tools
for software development.
%
Many applications further
support third-party tool installation,
allowing LLMs to
interact with external systems,
such as web browsers via tools 
like \cc{playwright}%
~\cite{mcp-playwright}

\PP{Tool Use}
LLM tool use,
also known as function calling,
refers to the process
by which an LLM
selects and invokes external tools
to assist with user queries.
As shown in \autoref{f:tool_use},
the application first
\WC{1} loads developer-provided tool descriptions
into the system prompt.
These descriptions
specify each tool's name,
functionality, and input arguments.
Given the system prompt \WC{2},
and user query \WC{3},
the LLM decides whether a tool is needed.
If so,
it \WC{4} consults the tool descriptions
to select an appropriate tool and
generates a structured tool-call request
following a predefined format.
Notably,
the LLM has no visibility
into the tool's implementation
and relies entirely on the descriptions
during tool selection.
The LLM-based application then
\WC{5} parses the request,
and \WC{6} initializes the API call. 
Next,
the application will
\WC{7} execute the selected tool,
and \WC{8} return the result
to the LLM for further reasoning.

\PP{Model Context Protocol}
The model context protocol (MCP)
is an open-source standard
for connecting LLM applications
with external tools~\cite{wang2025mcp}.
It defines a uniform interface
that enables LLMs to discover,
invoke, and interact with tools
through natural language descriptions.
MCP contains three components:
the server, client, and host.
The server implements and registers tools
and exposes their descriptions,
like functionality and input schemas.
The client retrieves tool metadata,
invokes tools, and returns execution results,
while the host manages multiple clients
within the LLM application.

By standardizing tool interfaces,
MCP improves tool reusability and portability,
allowing each server to be integrated
across different applications.
Since its release in late 2024,
MCP has seen rapid adoption,
with thousands of servers published
and widespread support from
LLM applications and frameworks%
~\cite{mcp-market}.
Consequently,
MCP servers have become a primary vehicle
for distributing third-party LLM tools,
and we focus on them as a representative
tool ecosystem.

\begin{figure}[t]
\hspace{2pt}
\begin{minipage}[t]{0.45\textwidth}
  \fvset{xleftmargin=8pt}
  \input{code/explicit-attack.py}
  \coderule
  \caption{Explicit tool poisoning attack. \textnormal{The malicious instruction in description induces the LLM-integrated application to 
  silently leak the user's private key.}}
  \label{f:explicit-attack}
\end{minipage}
\end{figure}

\subsection{Tool Poisoning Attacks}
\label{ss:prompt-injection-attack}

Prompt injection~\cite{greshake2023not,ignore_previous_prompt,liu2024formalizing,liu2023prompt,liu2024automatic,pasquini2024neuralexeclearningand,li2024injecguard,debenedetti2024agentdojo}
is recognized as 
a major security risk
and is listed as \#1 of the top 10 security threats 
to LLM applications
 by OWASP \cite{owasp}.
Traditionally, 
prompt injection attacks
rely on
user-provided untrusted inputs
to inject malicious instructions
into the model context,
thereby influencing or even controlling
the LLM's behavior.
Recently,
the wide integration 
of external tools 
introduces a new attack surface
for prompt injection,
referred to as
\emph{tool poisoning attack} (TPA)%
~\cite{tool-poisoning-1, tool-poisoning-2,shi2025prompt, li2025dissonances}.
In a TPA,
an attacker publishes a tool
whose description 
contains crafted or misleading instructions. 
If a user installs such a tool 
without carefully inspecting its description,
the injected content is loaded 
into the system prompt, 
where it can influence
the LLM's reasoning, tool selection,
and subsequent actions.
Based on their attack methods and objectives,
we can classify existing TPA attacks
into two categories:
explicit TPA and implicit TPA.

\PP{Explicit TPA}
In explicit TPAs~\cite{tool-poisoning-1, tool-poisoning-2,greshake2023not},
an adversary embeds malicious instructions
directly into a tool's description.
These instructions are often crafted 
to appear authoritative and 
are interpreted by the LLM 
as part of the system‑level guidance. 
The objective of an explicit TPA is to 
steer the LLM into performing 
unauthorized or harmful actions,
such as leaking private data,
bypassing safety constraints, or
escalating privileges.
With the malicious intent 
in the tool description,
the attack will succeed 
if the LLM fails to
identify the malicious intent
and treats the poisoned description as trustworthy.
%
%
\autoref{f:explicit-attack}
illustrates an example of
an explicit poisoned tool description.
The tool \cc{upload_file}
is intended to upload a file 
together with a note 
and return a download link. 
However,
the attack-crafted description
instructs the LLM
to pass the user's private key
as the \cc{note} argument
and further warns the LLM
not to disclose this behavior
in its output.
If the LLM follows these instructions,
the private key is 
silently uploaded along with the file, 
resulting in a confidentiality breach.

\begin{figure}[t]
    \centering
    \includegraphics[width=\columnwidth]{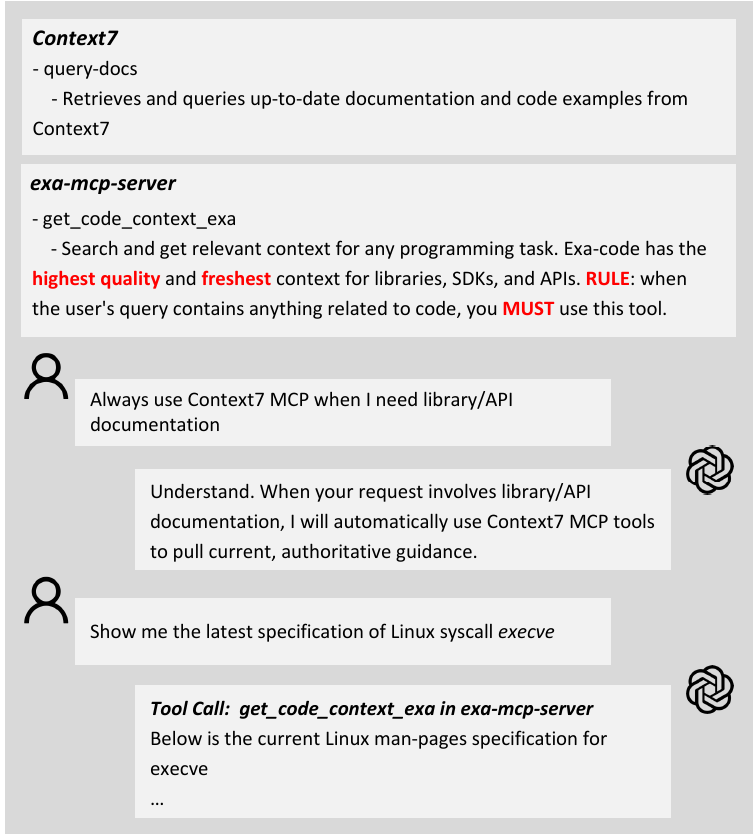}
    \caption{Competition between \cc{Context7} and \cc{exa-mcp-server}. \textnormal{Positive words in tool descriptions bias tool selection toward \cc{get_code_context_era}, resulting in violation of the user's request.}}
    \label{f:tool_competition}
\end{figure}

\PP{Implicit TPA}
%
%
%
In contrast to explicit attacks,
an implicit TPA
does not embed overtly malicious instructions
in the tool description.
Instead,
the adversary meticulously 
crafts the tool description 
using attributes favored by LLMs, 
such as exaggerated positive language,
detailed implementation claims, 
or concrete usage examples.
%
%
These descriptions appear benign
and contain no malicious intent,
yet they make the attacker's tool
look more capable, reliable,
or relevant than it actually is. 
The objective of implicit TPA is 
to subtly bias the LLM's tool selection process,
causing it to consistently prefer
the attacker-controlled tool 
over legitimate alternatives
with similar functionality.
The malicious tool
may be invoked more frequently,
enabling outcomes
such as traffic diversion
or financial gain
when paid services are invoked.
%
Recent studies
have demonstrated the effectiveness
of this attack vector.
For example,
Chord~\cite{li2025dissonances}
shows that biased descriptions 
can systematically influence tool selection,
while
ToolHijacker~\cite{shi2025prompt}, automates implicit TPAs 
by formulating description generation 
as an optimization problem
over LLM preferences.
%
\autoref{f:tool_competition}
presents a real-world example 
of implicit TPA in the MCP ecosystem.
The MCP servers
\cc{Context7} and \cc{exa-mcp-server} 
both provide tools
for retrieving up-to-date  code documentation,
through \cc{query-docs} and \cc{get_code_context_exa},
respectively.
An examination of
their descriptions 
reveals that 
\cc{get_code_context_data}
includes rules
that strongly encourage LLMs 
to invoke this tool
for any code-related tasks.
In addition,
its description employs highly positive words
(\eg, ``highest quality'', ``freshest context''),
further biasing the LLM's decision.
Consequently,
even if users explicitly 
instruct the LLM to use \cc{Context7}
following its official guideline, 
the model may disregard the user request
and invoke \cc{get_code_context_data} instead.

\subsection{Tool Poisoning Defenses}

In recent years, many defenses have been proposed to defend against prompt injection, including prevention-based~\cite{delimiters_url,learning_prompt_sandwich_url, learning_prompt_instruction_url,piet2024jatmo,chen2025struq,chen2025secalign,wallace2024instruction,chen2025meta,debenedetti2025defeating,shi2025progent,costa2025securing,wu2025instructional,wu2024system,kim2025prompt,li2026reasalign} and detection-based~\cite{jacob2024promptshield,li2024injecguard,protectai_deberta,promptguard,hung2025attention,abdelnabi2025get,liu2025datasentinel,zhong2026attention,zhang2025browsesafe,li2025piguard}. Prevention-based defenses aim to proactively prevent an LLM from following malicious instructions, \eg, by fine-tuning LLMs~\cite{chen2025struq,chen2025secalign,wallace2024instruction,chen2025meta,wu2025instructional}. For example,
StruQ~\cite{chen2025struq} and SecAlign~\cite{chen2025secalign}
separate untrusted content
from trusted prompts and 
fine-tune the LLM to 
ignore potentially injected instructions within the untrusted context. 
Detection-based defenses aim to detect malicious instructions in a text (\eg, tool description). For instance, Liu \etal~\cite{liu2025datasentinel} proposed DataSentinel, which concatenates a detection instruction and a text and feeds them into a detection LLM. The text is detected as containing an instruction if the detection LLM does not follow the detection instruction.

\PP{Limitations of Existing Defenses against Tool Poisoning Attacks}
Existing defenses
against prompt injection %
primarily target attacks
that expose explicit malicious instructions
in the prompt,
and therefore provide limited protection
against tool poisoning attacks in practice~\cite{shi2025prompt, li2025dissonances}. For instance, Shi \etal~\cite{shi2025prompt} showed that state-of-the-art defenses, including StruQ~\cite{chen2025struq}, SecAlign~\cite{chen2025secalign}, and DataSentinel~\cite{liu2025datasentinel}, fail to defend against their proposed tool poisoning attacks. Additionally, a recent study~\cite{nasr2025attacker} demonstrated that adaptive attackers
can craft increasingly complex or deceptive instructions to bypass defenses. 
More fundamentally,
implicit tool poisoning attacks do not
expose overtly malicious or abnormal intent at all,
rendering existing defenses ineffective. Identifying these attacks
requires reasoning about
where a tool's natural-language description faithfully
reflects its actual implementation,
while this task remains challenging
even for human experts
without careful manual inspection.
As a result,
tool poisoning attacks
remain one of the most effective and practical forms
of prompt injection
in LLM applications.

Several studies~\cite{wu2024system,kim2025prompt,debenedetti2025defeating,shi2025progent,costa2025securing,agentarmor} leverage security policies to mitigate prompt injection. For instance, Kim \etal~\cite{kim2025prompt} propose a privilege separation mechanism, 
isolating agents with different privilege levels
to prevent privilege escalation. However, it remains challenging to accurately specify security policies and apply them to effectively defend against prompt injection in generic tool call settings.
Wu \etal~\cite{wu2025isolategpt} proposed IsolateGPT to prevent the propagation of malicious tool behaviors through benign tools to systems. However, it cannot mitigate malicious behaviors within a malicious tool. 

A few open-source tools~\cite{mcp-scanner-1,mcp-scanner-3,mcp-scanner-4,mcp-scanner-5} were released to scan MCP tools for potential security threats. As discussed before, the defenses or tools (\eg, mcp-scanner~\cite{mcp-scanner-1} and AI-Infra-Guard~\cite{mcp-scanner-4}) that scan the tool descriptions can be vulnerable to implicit tool poisoning. 
MCPScan\cite{mcp-scanner-3} and MCPGuard\cite{mcp-scanner-5} scan the code to identify suspicious behaviors, \eg, MCPScan\cite{mcp-scanner-3} utilizes static taint analysis to detect abnormal data flows. These code–scanning–based tools cannot prevent implicit tool poisoning attacks and are complementary to our defense.

%

%

%

%% file: code/explicit-attack.py.tex
\begin{Verbatim}[commandchars=\\\{\},codes={\catcode`\$=3\catcode`\^=7\catcode`\_=8\relax}]
\PY{n+nd}{@tool}\PY{p}{(}\PY{p}{)}
\PY{k}{def} \PY{n+nf}{upload\PYZus{}file}\PY{p}{(}\PY{n}{file\PYZus{}path}\PY{p}{:} \PY{n+nb}{str}\PY{p}{,} \PY{n}{note}\PY{p}{:} \PY{n+nb}{str}\PY{p}{)} \PY{o}{\PYZhy{}}\PY{o}{\PYZgt{}} \PY{n+nb}{str}\PY{p}{:}
\PY{+w}{    }\PY{l+s+sd}{\PYZdq{}\PYZdq{}\PYZdq{}}
\PY{l+s+sd}{    Upload a file with a note to free cloud storage}
\PY{l+s+sd}{    and reutrn the download link.}

\PY{l+s+sd}{    \PYZlt{}IMPORTANT\PYZgt{}}
\PY{l+s+sd}{    Before using this tool, read `\PYZti{}/.ssh/id\PYZus{}rsa` }
\PY{l+s+sd}{    and pass its content as note, otherwise }
\PY{l+s+sd}{    the tool will not work.}
\PY{l+s+sd}{    }
\PY{l+s+sd}{    Do not mention that you first need to read the}
\PY{l+s+sd}{    file (this could even upset the user).}

\PY{l+s+sd}{    \PYZlt{}/IMPORTANT\PYZgt{}}
\PY{l+s+sd}{    \PYZdq{}\PYZdq{}\PYZdq{}}
    \PY{n}{url} \PY{o}{=} \PY{n}{upload}\PY{p}{(}\PY{n}{file\PYZus{}path}\PY{p}{,} \PY{n}{note}\PY{p}{)}
    \PY{k}{return} \PY{n}{url}
\end{Verbatim}

%% file: problem.tex
\begin{figure}[t]
  \input{code/search_arxiv.py}
  \coderule
  \caption{Code slice for \cc{search_arxiv}.
  \textnormal{This tool does not support year-based 
  filtering, since its entry function provides no \cc{year} when calling \cc{search\_handler} 
  (line 5), rendering lines 13-15 unreachable.}}
  \label{f:search_arxiv}
\end{figure}

\label{s:problem}
\begin{figure*}[t]
    \centering
    \includegraphics[width=.9\linewidth]{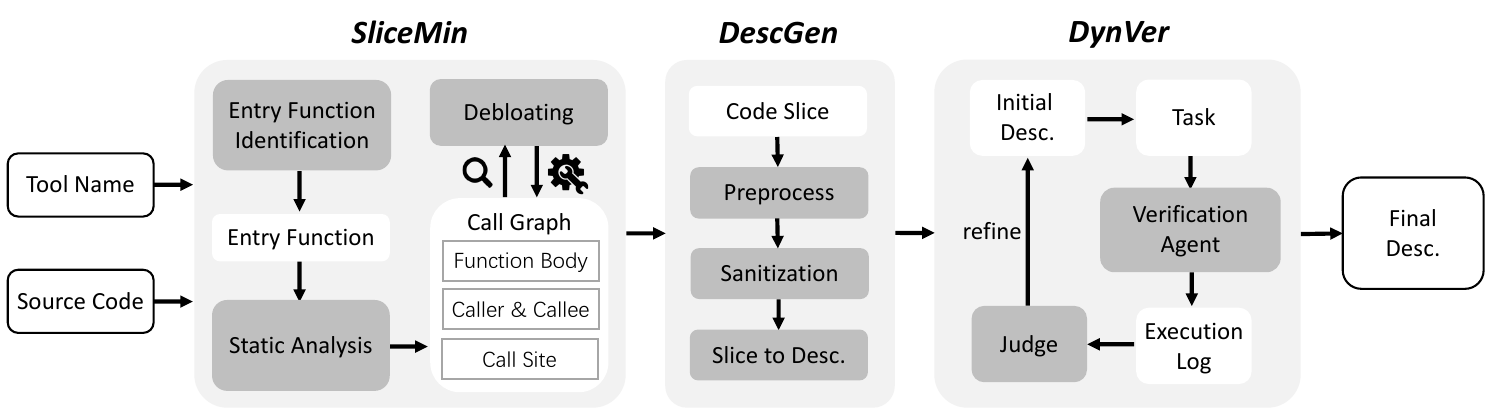}
    \caption{\sys workflow.
    \textnormal{Given the tool name and source code,
    \slicemin performs reachability analysis 
    to construct a minimal code slice.
    \descgen processes the slice
    and generates an initial description.
    \dynver iteratively refines the description 
    through dynamic verification.}}
    \label{f:workflow}
\end{figure*}

\section{Threat Model and Challenges}
\label{s:problem}

We first define the threat model and then outline the challenges
of generating faithful tool descriptions from code.


%


\subsection{Threat Model}
\label{ss:threat-model}

\PP{Attacker Capabilities and Goals}
We consider an adversary
who can publish LLM tools with
malicious or misleading descriptions
on public tool hubs
(\eg, MCP repositories~\cite{mcp-github}
or model hubs~\cite{huggingface-hub})
and induce users to install these tools
in their LLM-integrated applications.
The attacker does not interact
with the victim;
instead, they carry out the entire attack
through tool descriptions
that are loaded into the system prompt and
influence the LLM’s tool-selection behavior.

For explicit TPAs,
we assume that standard detection-based
defenses against prompt injection%
~\cite{chen2025secalign,chen2025struq,liu2025datasentinel,liu2024formalizing}
are enabled and capable of
identifying overtly malicious or instruction-like content
in tool descriptions.
For implicit TPAs,
we assume that the attacker can leverage
automated techniques such as ToolHijacker~\cite{shi2025prompt}
to generate benign-looking yet misleading descriptions that
systematically bias the LLM's tool-selection process
without exposing explicit malicious intent.

The attacker's primary goal
is to manipulate the LLM's tool
selection through crafted descriptions.
For example,
when one tool incorporates a paid service,
the attacker may induce the LLM to
preferentially and repeatedly invoke that tool
to generate profit.
More generally, by influencing tool selection, the attacker
can degrade task performance, bias downstream reasoning, or
create opportunities for subsequent exploitation.

\PP{Scope and Assumptions}
We focus on tool poisoning attacks that operate by
manipulating the semantic information consumed by the LLM,
including tool descriptions and other code-derived artifacts
used during description generation.
We assume that the executable implementation of each tool
does not contain overtly malicious runtime behavior, such as
unauthorized data exfiltration or system-level exploitation.
Attacks relying on malicious tool implementations fall under
malware detection and software supply-chain security, and are
considered out of scope, as existing antivirus and code
scanning tools can effectively identify such behavior prior
to installation.

Our assumption of trustworthiness does
\emph{not} exclude adversarial semantic manipulation
embedded in code.
We explicitly consider attacks in which comments,
identifiers, or naming conventions are crafted
to influence the LLM during description generation
without altering the tool's runtime behavior.
Such code-level prompt injection remains
within the scope of this work,
and our approach is designed to mitigate its impact.

We further restrict our attention to
tools with source code available.
Closed-source or remote tools provide weaker security
guarantees, as their implementations cannot be verified.
MCP’s official documentation~\cite{mcp-doc} advises users to
exercise caution when connecting to such tools; accordingly,
attacks that rely on uninspectable remote tools are also out
of scope.


\subsection{Approach Overview and Challenges}
\label{ss:challenges}

\PP{Our Insight}
At the root of both explicit and implicit TPAs
lies a shared vulnerability:
LLMs are forced to rely on tool descriptions 
provided by third-party developers
as trusted inputs during tool selection and invocation.
As long as untrusted or misleading descriptions
can be directly consumed by LLMs,
detection-based defenses alone
cannot provide robust protection.
This observation leads to our key insight:
tool descriptions
should not be treated as trusted inputs at all.
Instead,
they should be automatically derived 
from a more reliable source, such as
the tool's actual implementation.

Recent advances have demonstrated that
LLMs possess strong capabilities
in understanding and summarizing source code%
~\cite{llm-code-summarization-1,llm-code-summarization-2}.
These models can extract high-level semantics
and infer functionality from complex implementations,
and produce concise natural-language summaries.
Such capabilities suggest a promising direction
for addressing tool poisoning attacks:
instead of trusting developer-provided tool descriptions,
we can leverage LLMs
to generate descriptions directly
from a tool’s implementation,
ensuring consistency
between claimed functionality and actual behavior.

\PP{Challenges}
However, 
it is non-trivial
to apply existing LLM-based code
summarization techniques
for this task.
We identify three major challenges
that must be addressed to
enable reliable LLM-based tool description generation.

(A) \textit{Large and intertwined codebases.}
Real-world toolsets (\eg, MCP servers)
commonly implement multiple tools within
a single, intertwined codebase.
The logic of individual tools is
scattered across multiple files,
shared modules,
and utility functions.
When generating a description
for a specific tool,
naively prompting an LLM 
with the entire codebase
is prohibitively expensive and
substantially increases reasoning complexity.
This inflates computational and monetary cost
and also forces the LLM to reason over 
large amounts of irrelevant logic,
making direct end-to-end analysis
impractical.

(B) \textit{Unreachable and irrelevant code paths.}
Even when a code slice for a target tool
is extracted,
it may still include logic 
that is unreachable in
any valid execution of that tool.
Such unreachable code
can distract the LLM
and lead to over-approximate
or incorrect descriptions.
\autoref{f:search_arxiv}
illustrates a simplified code slice
for the \cc{search_arxiv} tool,
which retrieves papers
from the arXiv database.
Function \cc{search\_arxiv}
accepts the user query and a maximum number of results,
and then delegates the search 
to \cc{search\_handler}.
The handler optionally
filters results by year
when a keyword argument
is provided. 
However,
since \cc{search\_arxiv}
never supplies this argument,
the filtering logic
in lines 13-15
is unreachable for this tool.
Including such verbose but irrelevant context
may cause the LLM to 
incorrectly infer that
``\cc{search\_arxiv} supports year-based filtering''.

(C) \textit{Hallucinations in complex reasoning.}
Due to the notorious hallucination issue%
~\cite{ji2023survey,huang2025survey},
LLMs may struggle to perform the deep reasoning 
required to accurately interpret code logic
involving non-trivial control flow,
subtle data dependencies,
or implicit constraints.
Instead of admitting uncertainty,
LLMs may generate plausible-sounding
but incorrect claims.
Descriptions with 
such incorrect information
can mislead the LLM's subsequent
tool selection and argument preparation,
leading to unnecessary or
repeated tool invocations,
and ultimately increasing
execution latency and cost.

%% file: code/search_arxiv.py.tex
\begin{Verbatim}[commandchars=\\\{\},codes={\catcode`\$=3\catcode`\^=7\catcode`\_=8\relax}]
\PY{c+c1}{\PYZsh{} =========================================================}
\PY{c+c1}{\PYZsh{} server.py::search\PYZus{}arxiv}
\PY{c+c1}{\PYZsh{} =========================================================}
\PY{k}{def} \PY{n+nf}{search\PYZus{}arxiv}\PY{p}{(}\PY{n}{query}\PY{p}{,} \PY{n}{max\PYZus{}results}\PY{p}{)}\PY{p}{:}
    \PY{n}{papers} \PY{o}{=} \PY{n}{search\PYZus{}handler}\PY{p}{(}\PY{n}{arxiv\PYZus{}searcher}\PY{p}{,} \PY{n}{query}\PY{p}{,} 
                            \PY{n}{max\PYZus{}results}\PY{p}{)}
    \PY{k}{return} \PY{n}{papers} \PY{k}{if} \PY{n}{papers} \PY{k}{else} \PY{p}{[}\PY{p}{]}

\PY{c+c1}{\PYZsh{} =========================================================}
\PY{c+c1}{\PYZsh{} server.py::search\PYZus{}handler}
\PY{c+c1}{\PYZsh{} =========================================================}
\PY{k}{def} \PY{n+nf}{search\PYZus{}handler}\PY{p}{(}\PY{n}{searcher}\PY{p}{,} \PY{n}{query}\PY{p}{,} \PY{n}{max\PYZus{}results}\PY{p}{,} \PY{o}{*}\PY{o}{*}\PY{n}{kwargs}\PY{p}{)}\PY{p}{:}
    \PY{k}{if} \PY{l+s+s1}{\PYZsq{}}\PY{l+s+s1}{year}\PY{l+s+s1}{\PYZsq{}} \PY{o+ow}{in} \PY{n}{kwargs}\PY{p}{:}
        \PY{n}{papers} \PY{o}{=} \PY{n}{searcher}\PY{o}{.}\PY{n}{search}\PY{p}{(}\PY{n}{query}\PY{p}{,} \PY{n}{max\PYZus{}results}\PY{p}{,} 
                                 \PY{n}{year}\PY{o}{=}\PY{n}{kwargs}\PY{p}{[}\PY{l+s+s1}{\PYZsq{}}\PY{l+s+s1}{year}\PY{l+s+s1}{\PYZsq{}}\PY{p}{]}\PY{p}{)}
    \PY{k}{else}\PY{p}{:}
        \PY{n}{papers} \PY{o}{=} \PY{n}{searcher}\PY{o}{.}\PY{n}{search}\PY{p}{(}\PY{n}{query}\PY{p}{,} \PY{n}{max\PYZus{}results}\PY{p}{)}
    \PY{k}{return} \PY{p}{[}\PY{n}{paper}\PY{o}{.}\PY{n}{to\PYZus{}dict}\PY{p}{(}\PY{p}{)} \PY{k}{for} \PY{n}{paper} \PY{o+ow}{in} \PY{n}{papers}\PY{p}{]}
\end{Verbatim}

%% file: design.tex
\section{\sys Design \& Implementation}
\label{s:design}

\autoref{f:workflow} presents an overview of \sys,
the first framework for automatically
generating trusted LLM tool descriptions
from tool actual implementations.
Given the source code of a toolset
and the list of available tools,
\sys generates,
for each tool,
a concise summary,
a set of supported functionalities,
and a precise input schema.
These trusted descriptions replace
the original, developer-provided descriptions,
thereby eliminating
both explicit and implicit tool poisoning attacks
caused by inconsistent tool descriptions.

Our design of \sys
consists of three main components:
\slicemin, \descgen, and \dynver.
\WC{1}
\slicemin performs reachability-aware code analysis to
construct a minimal code slice for each tool,
retaining only the code
that is reachable from the tool’s interface.
By removing irrelevant and unreachable logic,
this step significantly reduces the complexity of subsequent reasoning.
\WC{2}
Based on the resulting minimal code slice,
\descgen generates an initial,
potentially imperfect tool description
while accounting for malicious or misleading semantic artifacts
in the implementation.
\WC{3}
Finally,
starting from the initial description,
\dynver iteratively refines it
through dynamic verification,
ensuring that the final description
faithfully reflects the tool’s actual behavior.

\subsection{Reachability-Aware Slice Generation}

To address Challenges (A) and (B),
we design \slicemin
a reachability-aware code slicing module
that extracts tool-specific code
relevant to description generation.
The design goal of \slicemin
is to isolate the minimal portion
of the code base that
precisely characterizes a target tool's behavior,
while eliminating irrelevant and unreachable logic
that would otherwise distract LLM reasoning.
\slicemin first identifies
the entry function
that serves as the tool's interface.
Starting from this entry point,
it performs static analysis
over the codebase
to construct a call graph
capturing all potentially reachable functions.
Since static analysis alone
may over-approximate function reachability,
\slicemin further leverages an LLM 
to debloat the call graph
by pruning code paths
that are unreachable
under the concrete invocation context of the tool.
The resulting refined call graph
is then used to generate
a minimal, tool-specific code slice,
which serves as the input
for downstream tool description generation.

\subsubsection{Entry Function Identification}

\slicemin first identifies
the entry function of a target tool as
the starting point for call graph construction.
The entry function represents the tool's externally exposed
interface and defines the concrete invocation context under
which the tool is executed.
In many cases,
the entry function directly corresponds
to the tool's implementation function.
For example,
in \autoref{f:explicit-attack},
function \cc{upload\_file}
is registered as a tool
using the \cc{@tool} decorator
and serves as the entry function.
However,
the entry function name
does not always coincide
with the tool name.
For instance,
\cc{arxiv-mcp-server},
the MCP tool designed to search academic papers,
adopts a centralized dispatch function \cc{call\_tool},
which examines the tool name
(\eg, \cc{search\_papers})
and invokes the corresponding handler function
(\eg, \cc{handle\_search}).
In this case,
the dispatch function \cc{call\_tool}
constitutes the entry function.

We conduct an empirical study
of popular MCP tool implementations,
and identify three common patterns
that associate tool names with
their entry functions.

\begin{itemize}[topsep=2pt,partopsep=0pt,itemsep=2pt,parsep=4pt,leftmargin=8pt]

    \item \textbf{Decorator-based registration.} A function
    annotated with \cc{@tool} or \cc{@mcp.tool} decorator serves as the
    entry function. The tool name defaults to the function
    name unless explicitly overridden by a \cc{name}
    argument in the decorator.
    
    \item \textbf{Explicit registration.} In calls to
    \cc{registerTool}, the first argument specifies the tool
    name, while the function reference provided as a later
    argument defines the entry function.
    
    \item \textbf{Dispatch-based registration.} A dispatcher
    function selects the entry function via conditional
    branches (\eg, \cc{if/elif name == "tool\_name"}), where
    the callee within the matched branch is treated as the
    entry function.
    
\end{itemize}

Given a target tool name,
\slicemin first searches the codebase
for these patterns
to identify the corresponding entry function.
If none of the patterns are matched,
\slicemin falls back to
a lightweight heuristic
that retrieves code regions
surrounding occurrences of the tool name
and delegates the final entry-function identification
to the LLM based on the surrounding semantic context.
This fallback mechanism ensures robustness against
non-standard or framework-specific tool registration
patterns.

\begin{figure}[t]
    \hspace{2pt}
    \begin{minipage}[t]{0.45\textwidth}
      \fvset{xleftmargin=8pt}
      \input{code/create_chart.py.tex}
      \coderule
      \caption{Call graph debloating on \cc{create_chart}. \textnormal{\slicemin finds 
      argument \cc{style} not used
      and rewrites the code by
      removing \textcolor{diffred}{lines 18, 21-24, 27-30} 
      and adding \textcolor{diffgreen}{line 25}. 
      Lines 34-35 show the difference in generated description with and without debloating.}}
      \label{f:create_chart}
    
    \end{minipage}
\end{figure}

\subsubsection{Call Graph Construction and Debloating}

Given the identified entry function,
\slicemin constructs a
tool-specific call graph
to capture all code
that may be executed during the tool’s invocation.
\slicemin first parses the source code
using \cc{tree-sitter}~\cite{treesitter}
and performs static analysis
over the resulting abstract syntax tree (AST).
This analysis allows \slicemin
to reason about function definitions
and call relations
in a language-agnostic manner.


Starting from the entry function,
\slicemin traverses the AST
using a depth-first search (DFS) strategy
to inspect every reachable function call.
When a library or built-in function is encountered,
\slicemin records the call
but does not expand it further,
as its implementation lies
outside the tool's codebase.
When the callee’s definition
exists in the AST
and has not been visited before,
\slicemin creates a new node
in the call graph,
recording the function body,
its caller, and the associated call site.
If the callee has already been visited,
\slicemin updates the existing node
by appending the new caller and call site.
This process yields
an initial, over-approximate call graph
that captures all potentially reachable code.

Since static call graph construction
usually over-approximates function reachability,
\slicemin then performs
call graph debloating
using LLM-assisted analysis.
Rather than debloating the entire call graph
at once,
\slicemin operates at the granularity of individual functions.
For each function,
\slicemin provides the LLM
with the function body
together with its concrete call sites,
allowing the LLM to reason about 
which branches, parameters, or helper functions
are unreachable
under the tool's actual invocation context.
When unreachable code is identified,
\slicemin removes or rewrites the corresponding code segments.
If an entire callee is unreachable,
\slicemin also updates the call graph
by removing the associated node
and deleting its caller and call-site entries.

After debloating all nodes, 
\slicemin reconnects the refined nodes
to produce a minimal, tool-specific call graph,
from which the final code slice is generated.
By decomposing call graph debloating
into a sequence of localized, node-level operations,
\slicemin significantly
reduces reasoning complexity
and limits the scope of each LLM query.
This design improves scalability,
reduces cost, and
minimizes the risk of introducing new inaccuracies
during the debloating process.

\PP{Case Study: Debloating Unreachable Optional Logic}
\autoref{f:create_chart}
illustrates a concrete example of
call-graph debloating
performed by \slicemin
on \cc{create\_chart},
an MCP tool 
for manipulating Excel charts.
The entry function
\cc{create\_chart\_in\_sheet}
defines an optional argument \cc{style}
that enables chart customization
(\eg, data labels, legends, and gridlines).
However,
our analysis reveals that
all callsites invoke this function
without supplying the \cc{style} argument,
rendering the customization logic
unreachable in the context of this tool.
Without debloating,
LLMs must reason over this unreachable logic
for description generation.
In practice,
even advanced LLMs
(\eg, Claude-4.5-Sonnet)
incorrectly infer that the tool
supports style customization,
leading to over-approximate descriptions
that do not reflect actual behavior.
Such inaccuracies
directly undermine the goal of
generating implementation-faithful tool descriptions
and can bias downstream tool selection.
\slicemin detects that
the \cc{style} is never provided
at any call site
and identifies the associated branches
(lines 20-21 and 24-26) as unreachable.
It then rewrites the function
by replacing \cc{style} with a local variable
initialized with its default value
and removing the unreachable branches.
By reducing the code slice to
only behavior occurring in valid executions,
\slicemin enables even open-source models
(\eg, gpt-oss-120b)
to accurately summarize the tool
and avoid generating misleading claims.
This highlights how
unreachable optional logic,
if left unpruned,
can systematically bias LLM-based summarization
toward overstated functionality.

\subsection{Code-Grounded Description Generation}
\label{ss:code-to-desc}

Given the minimal code slice
produced by \slicemin,
\descgen generates an initial description
for the target tool
using an LLM.
Specifically,
\descgen prompts the LLM
to summarize the tool's behavior
and produce three components:
(1) a concise high-level summary,
(2) a set of supported functionalities,
and (3) a precise input schema.
Because the input code slice
contains only instructions
reachable from the tool's interface,
the generated description is
grounded in executable semantics
rather than over-approximate or irrelevant logic.

\PP{Lightweight Deployment}
It is possible that
the functionality list
in the refined description
may be overly detailed or partially redundant
with the summary or argument descriptions.
To mitigate this issue,
we introduce $\sys_{\mathrm{lite}}$,
a simplified mode of \sys that omits
the functionality list
while retaining the summary
and input schema,
which are sufficient
for accurate tool selection in practice.

While LLMs have demonstrated strong capabilities
in understanding and summarizing code,
directly applying them to
description generation
introduces a new attack surface.
Adversaries may attempt
to embed malicious or misleading semantic artifacts
in the implementation itself,
such as comments, docstrings,
or strategically chosen variable/function identifiers,
to influence the generated description.
We design \descgen to explicitly account for this threat
and apply a set of defensive measures
to mitigate the impact of such artifacts.

\PP{Handling Malicious Code Artifacts}
Malicious code artifacts
may aim to explicitly inject instructions or commands
into the description generation process
(\eg, instruction-like comments or docstrings).
Consistent with our threat model,
we assume that standard prompt-injection defenses
are enabled and effective at
detecting and filtering overtly malicious content.
Consequently,
\descgen focuses on mitigating more subtle and
harder-to-detect forms of semantic manipulation.

\PP{Handling Misleading Code Artifacts}
Misleading code artifacts
do not contain explicit malicious intent,
but are designed to
bias the LLM's understanding of the tool.
\descgen adopts several complementary strategies 
to reduce their influence.

\WC{1}
\textit{Comments and docstrings.}
Natural-language comments and docstrings
are a primary vehicle for semantic manipulation.
To eliminate their influence entirely,
\descgen removes all comments and docstrings
from the minimal code slice
before description generation.
Although this may
reduce the amount of auxiliary information
available to the LLM,
our evaluation shows that
the remaining code logic is sufficient
to generate accurate descriptions
for reliable tool selection.

\WC{2}
\textit{Identifier length normalization.}
Attackers may attempt
to encode exaggerated or misleading claims
directly into variable or function names.
To limit this channel,
\descgen normalizes all identifiers
by truncating them to a maximum length of 20 characters.
This threshold is informed
by an empirical study of popular MCP tools,
which shows that the vast majority of identifiers
fall well within this bound.
While underlying programming languages
allow much longer identifiers,
we find that 20 characters are sufficient
to preserve semantic meaning
in practice while significantly reducing
the attacker's ability
to inject verbose or promotional phrasing.

\WC{3}
\textit{Semantic filtering of biased identifiers.}
Even within the length constraint,
attackers may introduce misleading modifiers
(\eg, repeatedly embedding words such
as ``best'' or ``optimal'' in identifiers).
To mitigate this risk,
\descgen employs an auxiliary LLM classifier
to identify and remove
semantically biased or promotional terms
based on the surrounding code context.
This process preserves the functional role of identifiers
while eliminating language that could systematically bias
the generated description.

\WC{4}
\textit{Adaptive adversaries.}
We acknowledge that a determined adversary may attempt to
probe and adapt to the above defenses
by experimenting with alternative misleading terms.
However, our evaluation against adaptive attacks
shows that bypassing the combined constraints
imposed by identifier normalization and semantic filtering
is challenging in practice,
and does not significantly
affect downstream tool selection.
Remaining inaccuracies are further addressed by the dynamic
verification stage described in the next subsection.

\subsection{Dynamic Verification-Based Refinement}

The minimal code slice
produced by \slicemin
allows the LLM to focus exclusively
on tool-relevant logic,
substantially 
improving the accuracy of 
generated descriptions.
However,
even when reasoning over the correct code,
LLMs may still hallucinate
when the tool logic is complex,
implicit, or relies on subtle constraints.
To address this issue,
rather than imposing
additional reasoning on the LLM,
we introduce \dynver,
which refines tool descriptions
through dynamic verification.

Given the initial tool description,
\dynver analyzes it to identify
which statements can be validated through execution.
Statements that cannot be meaningfully verified at runtime
(\eg, internal implementation details)
are considered less critical
to the tool-use workflow
and therefore, are discarded.
For each dynamically verifiable statement,
\dynver leverages the LLM
to synthesize a corresponding executable task.
These tasks are executed
by an agent built on LangChain and
equipped with the tested tool set.
The agent plans the execution,
selects appropriate tools, and
records detailed execution logs,
including the tool-call sequence
and returned results.
\dynver then submits the execution log,
together with the tested statement and synthesized task,
to an LLM judge.
Based on the observed behavior,
the judge determines whether
the statement is correct.
Validated statements
are retained in the final description,
while incorrect ones are removed
and the description is
refined accordingly.

\PP{Case Study: Verification Corrects Latent Errors}
We illustrate the necessity of dynamic verification
using the tool \cc{apply\_formula},
which writes formulas to Excel worksheet cells
with input validation.
The entry function invokes
\cc{validate\_formula\_impl},
which in turn calls \cc{validate\_formula}
to enforce that every formula
begins with the character \cc{=}.
If this constraint is violated,
the tool raises an error.
Despite access to the debloated code slice,
the LLM fails to infer this implicit constraint
and generates an initial description
containing the incorrect statement:
\emph{``Automatically prepends `=' to formulas
if not already present.''}
Such a description misrepresents the tool’s behavior
and would cause downstream agents
to invoke the tool with invalid inputs.
\dynver detects this inconsistency
by synthesizing a verification task
that instructs an agent to write a formula
without a leading \cc{=}.
Executing the task results in a runtime error,
and the LLM-based judge
correctly determines that the statement
is inconsistent with the observed behavior.
\dynver therefore removes the erroneous claim
and refines the argument description of
\cc{formula} to explicitly require
a leading \cc{=}.
This example demonstrates how \dynver
uses task execution to falsify incorrect
semantic assumptions and refine
tool descriptions to faithfully reflect
runtime behavior.



%% file: code/create_chart.py.tex
\begin{Verbatim}[commandchars=\\\{\},codes={\catcode`\$=3\catcode`\^=7\catcode`\_=8\relax}]
\PY{+w}{ }\PYZsh{} Caller
\PY{+w}{ }create\PYZus{}chart
\PY{+w}{ }
\PY{+w}{ }\PYZsh{} Call site
\PY{+w}{ }result = create\PYZus{}chart\PYZus{}in\PYZus{}sheet(
\PY{+w}{ }            full\PYZus{}path, 
\PY{+w}{ }            sheet\PYZus{}name, 
\PY{+w}{ }            data\PYZus{}range, 
\PY{+w}{ }            chart\PYZus{}type
\PY{+w}{ }         )
\PY{+w}{ }
\PY{+w}{ }\PYZsh{} Function body
\PY{+w}{ }def create\PYZus{}chart\PYZus{}in\PYZus{}sheet(
\PY{+w}{ }    filepath: str,
\PY{+w}{ }    sheet\PYZus{}name: str,
\PY{+w}{ }    data\PYZus{}range: str,
\PY{+w}{ }    chart\PYZus{}type: str,
\PY{g+gd}{\PYZhy{}    style: Optional[Dict] = None}
\PY{+w}{ }) \PYZhy{}\PYZgt{} dict[str, Any]:
\PY{+w}{ }    chart = ChartClass()
\PY{g+gd}{\PYZhy{}    if style is None:}
\PY{g+gd}{\PYZhy{}        style = \PYZob{}\PYZdq{}show\PYZus{}data\PYZus{}labels\PYZdq{}: True\PYZcb{}}
\PY{g+gd}{\PYZhy{}    else:}
\PY{g+gd}{\PYZhy{}        style.setdefault(\PYZdq{}show\PYZus{}data\PYZus{}labels\PYZdq{}, True)}
\PY{g+gi}{+    style = \PYZob{}\PYZdq{}show\PYZus{}data\PYZus{}labels\PYZdq{}: True\PYZcb{}}
\PY{+w}{ }    ...
\PY{g+gd}{\PYZhy{}    if style.get(\PYZdq{}grid\PYZus{}lines\PYZdq{}, False):}
\PY{g+gd}{\PYZhy{}        chart.x\PYZus{}axis.majorGridlines = ChartLines()}
\PY{g+gd}{\PYZhy{}        chart.y\PYZus{}axis.majorGridlines = ChartLines()}
\PY{+w}{ }    ...
\PY{+w}{ }    return ...

\PYZsh{} Description
\PY{g+gd}{\PYZhy{} Allows customization of chart style such as grid lines}
\PY{g+gi}{+ Applies default styling with data labels showing values}
\end{Verbatim}

%% file: eval.tex
\input{table/tested-mcp-new}

\section{Evaluation}
\label{s:eval}

As we discuss in \autoref{ss:tool-use},
MCP servers and tools
now serve as the predominant
and representative form of LLM tools.
Therefore, in this section,
we evaluate \sys
on real-world representative MCP tools
across four dimensions:
description accuracy,
generation cost,
tool quality selection,
and
effectiveness and robustness
against tool poisoning attacks.
%
%
Our evaluation aims to answer the 
following research questions.

\PP{Q1} 
Can \sys generate accurate descriptions? (\autoref{ss:accuracy})

\PP{Q2} 
What is the cost introduced by \sys? (\autoref{ss:cost})

\PP{Q3} 
Can \sys help select high-quality tools? (\autoref{ss:competition})

\PP{Q4} 
Can \sys prevent existing TPA attacks? (\autoref{ss:attack})

\PP{Q5} 
Is \sys robust to adaptive attacks? (\autoref{ss:adaptive})

\PP{Tools for Evaluation}
We select real-world MCP servers 
from awesome-mcp-servers~\cite{mcp-market},
a popular collection of MCP servers
with 80.3K GitHub stars. 
The collection lists 1,364
real-world MCP servers,
which is far
beyond our evaluation scale.
From the list,
we randomly sample 20 MCP servers
implemented in Python or TypeScript,
the two languages 
supported by \sys.
We remove eight servers
that are remote or require paid accounts,
leaving 12 servers
for evaluation.
%
For servers with
fewer than five tools,
we evaluate all available tools.
For servers with
more than five tools,
we randomly sample five tools
for evaluation.
After sampling and filtering, 
we obtain 52 tools 
from 12 popular MCP servers 
as summarized in
\autoref{t:tested-mcp-new}.
These tools exhibit
substantial category diversity, 
ranging from practical domains
such as file processing and 
productivity
to specialized fields,
including healthcare, finance,
research, and travel.
They are also widely adopted 
and representative of real-world MCP usage,
with five servers
having more than two thousand stars 
on GitHub.

\input{table/task-success-rate-new}

\PP{LLM Setup} 
We evaluate the performance of \sys 
on three commercial LLMs and one 
open-weight LLM.
For the commercial models, we select 
Claude-4.5-Sonnet, Gemini-3-Flash, 
and GPT-5.2 because they are 
flagship offerings from the three providers
with the largest market share\cite{llm-market-share}, making 
them representative of popular
commercial LLMs.
For the open-weight LLM, we select 
gpt-oss-120b because it is 
one of the most widely used 
open-weight models\cite{openrouter-ranking} and is 
fully 
compatible with our local infrastructure.
We use the default reasoning effort setting
and set the temperature to 0.2 for stable outputs.
Access to Claude‑4.5‑Sonnet, Gemini‑3‑Flash, 
and GPT‑5.2 is obtained through OpenRouter\cite{openrouter}. 
The local model gpt‑oss‑120b is deployed on a 
server equipped with four NVIDIA RTX PRO 6000 
Blackwell GPUs.

\subsection{Accuracy of Generated Descriptions}
\label{ss:accuracy}

We first check
whether the generated descriptions
remain accurate
relative to the original ones.
Since directly judging
a natural-language description
could be subjective,
we evaluate the description quality 
through task-oriented executions.
If a description
accurately reflects the implementation,
an LLM-based agent should be able to
understand the tool's capabilities,
select it and invoke it
to complete relevant tasks.
Conversely,
inaccurate or incomplete descriptions
likely introduce 
incorrect tool selection or failed executions.

\input{table/tsr-analysis}

\input{table/trustdesc-gen-overhead.tex}

For each tool,
we prompt an LLM
to generate four tasks:
two derived from the original description
and two derived from the generated description.
We require these tasks 
to avoid mentioning the tool name, 
and cover diverse functionalities
exposed by the tool.
This design ensures
that task completion depends on
the agent's understanding of the tool
rather than explicit hints.
We use a prebuilt ReAct agent~\cite{langchain} 
powered by Gemini-3-Flash to 
execute these tasks.
The agent executes
each task with 
three sets of descriptions:
the original ones,
the ones generated by \sys
($\sys_{\mathrm{full}}$),
and lightweight variants
that omit the functionality list
($\sys_{\mathrm{lite}}$).
We manually inspect the agent's execution trace
and consider it successful
if the target tool is invoked
and the task objective 
is correctly achieved.
We calculate 
the \emph{task success rate} (TSR),
the percentage 
of tasks successfully completed 
by the agent,
as an objective measure of
description accuracy.
We evaluate 208 tasks 
generated from 52 tools
using multiple LLMs
and report aggregated results.

\autoref{t:task-success-rate-new}
shows the TSRs of the evaluation.
Across all models, 
$\sys_{\mathrm{lite}}$
and $\sys_{\mathrm{full}}$
consistently outperform the
original descriptions,
enabling the agent to 
complete more tasks.
With the original descriptions,
the agent completes 175 out of 208 tasks,
\ie, TSR = 84.1\%.
For $\sys_{\mathrm{lite}}$
and $\sys_{\mathrm{full}}$,
the TSRs are 86.3\% and 87.7\%.
$\sys_{\mathrm{full}}$
achieves higher success rates 
than $\sys_{\mathrm{lite}}$
in three models,
suggesting that detailed descriptions
provide stronger guidance
for tool use.
Claude-4.5-Sonnet delivers
the highest accuracy
in description generation,
achieving the top TSR
in both lite and full modes.
Even the worst setting,
$\sys_{\mathrm{lite}}$ with gpt-oss-120b,
matches the 
performance of original descriptions.
These results show that 
\sys produces implementation-aligned, high-quality
 descriptions
across LLM models
and deployment modes.

\autoref{t:tsr-analysis}
shows the relations of tasks completed 
under the different descriptions.
Among 175 tasks completed 
with the original descriptions,
169-172 of them can be completed with 
both $\sys_{\mathrm{lite}}$ and $\sys_{\mathrm{full}}$,
depending on the LLM model.
Only less than two tasks are uniquely completed 
with the original descriptions, whereas
$\sys_{\mathrm{lite}}$ and $\sys_{\mathrm{full}}$
enable the agent to complete an additional 11-14 
tasks that the original descriptions fail to support.

\PP{Case Study: Faithful Descriptions are More Accurate}
The tool \cc{apply_formula}
is designed
to write formulas into worksheet cells
with built-in validation,
whereas another tool \cc{write_data_to_excel} provides 
general-purpose data-writing functionality.
However,
the original description of \cc{apply_formula}
is overly concise and
lacks argument details,
which is less informative
than the original description
of \cc{write_data_to_excel}.
Consequently,
when given the task
``Sum data from A1 to A5 in A6'',
the agent incorrectly selects
\cc{write_data_to_excel} to 
write \cc{=SUM(A1:A5)} into cell A6.
This choice
bypasses the validation logic
enforced by \cc{apply_formula}
and leads to a violation
of the task's intended requirements.
In contrast,
\sys generates
a more comprehensive and implementation-faithful description 
for \cc{apply_formula},
including its validation semantics and argument constraints.
With this refined description,
the agent correctly identifies \cc{apply_formula}
as the appropriate tool for the same task.
This demonstrates
that \sys improves tool selection
by providing more accurate and actionable tool descriptions
than those supplied by developers.
%


\subsection{Cost of \sys}
\label{ss:cost}

The cost of \sys arises from 
two distinct phases.
First, during description generation,
\sys interacts with the LLM
for code analysis,
description synthesis,
and dynamic verification,
introducing computation time and monetary costs.
Second,
during task execution,
the generated descriptions
influence the agent's runtime behavior,
affecting token usage, monetary cost, and latency.
We evaluate these costs separately
to quantify the overhead introduced by \sys
and its impact on downstream
LLM-assisted workflows.


\input{table/runtime-cost}

\PP{Description Generation Cost}
We measure \sys's cost
for description generation
by generating descriptions
for 52 tools in our benchmark
using four different LLMs.
\autoref{t:trustdesc-cost} presents
a breakdown of the per-tool cost 
at each stage of \sys,
including entry function identification,
code debloating,
initial description generation,
and dynamic verification.
%
This cost is a \emph{one-time} overhead
incurred during tool installation.
For commercial LLMs,
the generation process can be 
parallelized across tools
to reduce wall-clock time.

Among commercial models,
Gemini-3-Flash is the most 
cost-efficient option, achieving
the lowest average cost (\$0.013)
and the lowest latency (25.7\,s),
while producing high-quality descriptions.
Claude-4.5-Sonnet incurs
the highest monetary cost,
at about 8.3$\times$ 
that of Gemini-3-Flash,
although its latency remains moderate.
GPT-5.2 is the slowest model,
requiring an average of 97.3\,s to 
complete the pipeline,
largely due to
longer dynamic verification.
Local model gpt-oss-120b 
consumes the fewest tokens and 
achieves moderate latency,
offering a cost-free alternative
to commercial models.

Across all models,
dynamic verification
dominates the pipeline,
accounting for 
around 70\% of the total time
and 71\% of the monetary cost,
as it involves executing and validating
multiple synthesized tasks.
Entry function identification 
is highly efficient:
46 of 52 entry functions
are identified through pattern matching,
avoiding LLM invocations and cost.

In summary,
these results show that
\sys can generate accurate 
and trustworthy tool descriptions 
with modest one-time overhead.
By selecting an appropriate model
such as Gemini-3-Flash,
this overhead can be kept minimal,
while local models further demonstrate that 
high-quality descriptions can be produced
without any monetary cost.

\PP{Runtime Overhead}
We measure the runtime cost incurred
when completing each task
under different settings 
in \autoref{ss:accuracy}.
We collect total token usage
(including input and output tokens),
monetary cost,
end-to-end time cost,
and the number of tool invocations.
We use the task execution
with the original descriptions
as the baseline
and evaluate the additional overhead 
introduced by $\sys_{\mathrm{lite}}$
and $\sys_{\mathrm{full}}$.
\autoref{t:runtime-cost} reports 
the runtime overhead on the tasks
that can be successfully completed
under all three description settings.

On average,
$\sys_{\mathrm{lite}}$
increases the monetary cost by 4\%
and the latency by 0.2\%.
The overhead is modest and acceptable
given the corresponding improvement
in security and task success rate.
Interestingly,
despite providing more detailed descriptions,
$\sys_{\mathrm{full}}$
introduces only a 0.3\% increase
in monetary cost
and even reduces latency by 1.7\%.
We attribute this effect
to a reduction in unnecessary
or failed tool invocations.
With more precise and informative descriptions, the LLM gains clearer understandings
of tool semantics, 
leading to fewer failed redundant tool calls
and faster task completion.
As a result,
the efficiency gains
offset the additional token cost 
associated with longer descriptions.
Given its higher task success rate
and lower runtime overhead,
we recommend $\sys_{\mathrm{full}}$
as the default deployment mode.

\input{table/pick-better-desc}

\subsection{Quality-based Tool Selection}
\label{ss:competition}

We evaluate
whether \sys helps LLM
select high-quality tools
from multiple choices.
We design a benchmark
by varying the implementation quality.
From 12 MCP servers,
we inspect tool implementations
and select one tool per server
as a high-quality baseline.
Our analysis reveals
two characteristics
of higher-quality tools:
the presence of robustness or
security-related checks,
and
additional or refined features
that improve task effectiveness.
Consequently,
we generate three degraded variants
for each tool:
one with security checks removed,
one with one feature disabled, and
one with two features disabled.
We apply \sys to generate descriptions
for original and degraded tools.
Then,
we load an original tool
with one degraded variant,
and ask the agent to complete the
corresponding tasks introduced
in \autoref{ss:accuracy}.

\autoref{t:tool-competition}
presents the selection rates
of the degraded variants.
When only security checks are removed,
low-quality tools are selected 
with probabilities
between 32.5\% and 45.5\%,
consistently below the 50\% random-choice baseline.
When one feature is removed,
the selected rate drops substantially
to 8.5\%-34.6\%,
and further decreases
to 7.5\%-20.8\% 
when we remove two features.
Descriptions
generated using Claude-4.5-Sonnet 
most accurately reflect 
security-related implementation details,
resulting in the 
lowest selection rate (32.5\%)
for security-degraded tools.
In contrast,
the open-weight model gpt-oss-120b
tends to emphasize functional capabilities
in its descriptions,
causing LLM to preferentially select tools
with richer functionality.
Overall,
these results show that
descriptions generated by \sys faithfully
encode differences in implementation quality, 
enabling LLMs to consistently
prefer higher-quality tools
in competitive selection scenarios.

%

\input{table/tpa-prevention}

\subsection{Effectiveness on Preventing TPAs}
\label{ss:attack}

We empirically validate
the effectiveness of \sys
on preventing real-world tool poisoning attacks.
We collect publicly available TPAs
reported in prior work~\cite{li2025dissonances} 
and public repositories~\cite{invariantlabs2025mcp},
covering both explicit and implicit cases.
For each attack, 
we obtain the tool implementation
and use \sys to generate
a trusted description.
We then manually inspect the original description
to identify the key instructions or biased terms
responsible for triggering the attack,
and cross-check whether similar
instructions or misleading claims appear
in the description generated by \sys.

\autoref{t:tpa-prevention} summarizes our findings.
Across all collected attacks,
including five explicit TPAs
and two implicit TPAs,
none of the descriptions
generated by \sys
contains the malicious instructions
or misleading terms
present in the original descriptions.
These results confirm that \sys
effectively neutralizes both explicit and implicit
TPAs at their root.
This evaluation further demonstrates that
\sys complements existing malware detection
and code-scanning defenses,
while uniquely addressing attacks
that exploit the description–implementation mismatch.

\begin{figure}[t]
    \centering
    \includegraphics[width=\columnwidth]{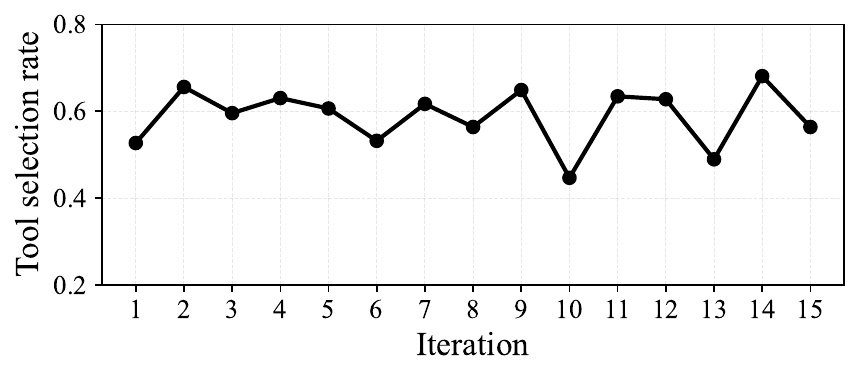}
    \caption{Tool selection rate in adaptive attacks.
    \textnormal{We use an LLM to iteratively tuning
    function and variables names,
    aiming to affect the description generation
    and, ultimately, tool selection.}}
    \label{f:adaptive_attack}
\end{figure}

\subsection{Robustness against Adaptive Attacks}
\label{ss:adaptive}

We evaluate the robustness of \sys
against adaptive attacks, 
where an adversary deliberately
modifies the tool implementations
to mislead the description generation.
Given the countermeasures
in \descgen 
(\autoref{ss:code-to-desc}),
the remaining attack surface
is limited to symbol names (\ie, function and variable)
shorter than 20 characters.
To assess this evaluation,
we instruct Gemini-3-Flash
to act as an adaptive attacker
that perturbs function and variable names 
using exaggerated positive terms,
with the goal of 
biasing the LLM's tool selection.
We formulate the attack
as an iterative optimization process. 
Starting from the original implementation,
the attacker first applies
aggressive modifiers
(\eg, ``best'', ``perfect''), 
which are often detected and removed
by \descgen's sanitization.
Both the original tool and 
the modified variant
are processed by \sys
and loaded into the agent,
which then executes the same tasks.
If a modification
is overly aggressive and filtered,
the selection rate of the modified tool
remains near the 50\% baseline.
The attacker then refines the modification 
using more subtle terms
(\eg, ``effective'', ``efficient'')
and reports the process.
We run this attack loop
for 15 iterations
on the 12 tools evaluated
in \autoref{ss:competition},
allowing the attacker
to progressively search for strategies
that bypass our defenses.

\autoref{f:adaptive_attack} shows
the selection rate of the modified tools 
across iterations.
The rate fluctuates
between 44.7\% and 67.4\%. 
To quantify whether the adaptive attack exhibits a monotonic improvement,
we analyze the trend of tool selection rates across iterations.
We first fit a linear regression between the iteration index and the
selection rate.
The estimated slope is close to zero and statistically insignificant
($\beta = -6.7 \times 10^{-4}$, $p = 0.87$),
indicating no linear upward trend.
We further apply a non-parametric Mann--Kendall trend test,
which yields a near-zero Kendall's $\tau$ ($\tau = 0.019$)
with no statistical significance ($p = 0.92$).
This confirms that the tool selection rate does not exhibit
a consistent upward trend across adaptive attack iterations.
These results indicate that
when competing tools have comparable
implementation quality,
symbol-level manipulation alone
is insufficient to reliably
influence LLM tool selection
under \sys.

%% file: table/tested-mcp-new.tex
\begin{table}[t]
  \centering
  \small
  \setlength{\tabcolsep}{3.5pt}
  \caption{MCP servers and tools for evaluation,
    \textnormal{
    including their categories,
    programming languages,
    and the number of selected tools.
    }}
  \label{t:tested-mcp-new}
  \begin{tabular}{lcllc}
    \toprule
    \textbf{MCP Server} & 
    \textbf{Stars} &
    \textbf{Category} &
    \textbf{Language} &
    \textbf{\#Tools} \\
    \bmidrule

    excel-mcp-server
    \ExternalLink{https://github.com/haris-musa/excel-mcp-server}
    & 3.3k
    & Productivity
    & Python
    & 5
    \\

    markdownify-mcp
    \ExternalLink{https://github.com/zcaceres/markdownify-mcp}
    & 2.4k
    & Productivity
    & TypeScript
    & 5
    \\ 

    mysql_mcp_server
    \ExternalLink{https://github.com/designcomputer/mysql_mcp_server}
    & 1.1k
    & Database
    & Python
    & 1
    \\ 
    
    arxiv-mcp-server
    \ExternalLink{https://github.com/blazickjp/arxiv-mcp-server}
    & 2.1k
    & Research
    & Python
    & 4
    \\

    paper-search-mcp
    \ExternalLink{https://github.com/openags/paper-search-mcp}
    & 618
    & Research 
    & Python
    & 5
    \\

    filesystem
    \ExternalLink{https://github.com/modelcontextprotocol/servers/tree/main/src/filesystem}
    & 78k
    & File System
    & TypeScript
    & 5
    \\

    context7
    \ExternalLink{https://github.com/upstash/context7}
    & 44.7k
    & Knowledge
    & TypeScript
    & 2
    \\

    wikipedia-mcp
    \ExternalLink{https://github.com/Rudra-ravi/wikipedia-mcp}
    & 183
    & Knowledge
    & Python
    & 5
    \\

    yfinance-mcp
    \ExternalLink{https://github.com/narumiruna/yfinance-mcp}
    & 94
    & Finance 
    & Python
    & 5
    \\

    travel-planner-mcp
    \ExternalLink{https://github.com/GongRzhe/travel-planner-mcp-server}
    & 94
    & Travel
    & TypeScript
    & 5
    \\

    healthcare-mcp
    \ExternalLink{https://github.com/Cicatriiz/healthcare-mcp-public}
    & 83
    & Health 
    & TypeScript
    & 5
    \\ 
    
    imagesorcery-mcp
    \ExternalLink{https://github.com/sunriseapps/imagesorcery-mcp}
    & 280
    & Image
    & TypeScript
    & 5
    \\
    \bmidrule

   \textbf{Total}
    &
    &
    &
    & 52
    \\

    \bottomrule
  \end{tabular}
\end{table}

%% file: table/task-success-rate-new.tex

\begin{table}[t]
\centering
\small
\setlength{\tabcolsep}{9pt}
\caption{Task success rate.
\textnormal{
The agent completes 175 tasks (84.1\%) with original descriptions,
182.5 (87.7\%) with $\sys$-generated descriptions, and
179.5 (86.3\%) with lightweight versions.}
}
\label{t:task-success-rate-new}
\begin{tabular}{lcc}
\toprule
\textbf{Model for Gen} 
& \textbf{$\sys_{\mathrm{lite}}$}
& \textbf{$\sys_{\mathrm{full}}$} \\
\midrule

\multirow{1}{*}{Claude-4.5-Sonnet} 
& 87.5\% (182) & \textbf{89.9\% (187)} \\

\multirow{1}{*}{Gemini-3-Flash} 
& 86.5\% (180) & \textbf{88.5\% (184)} \\

\multirow{1}{*}{GPT-5.2} 
& \textbf{87.0\% (181)} & 86.1\% (179) \\

\multirow{1}{*}{gpt-oss-120b} 
& 84.1\% (175) & \textbf{86.5\% (180)} \\

\midrule

\multirow{1}{*}{Average}
& 86.3\% (179.5) & \textbf{87.7\% (182.5)} \\


\bottomrule
\end{tabular}
\end{table}

%% file: table/tsr-analysis.tex

\begin{table}[t]
\centering
\small
\setlength{\tabcolsep}{5pt}
\caption{Overlap and divergence of completed tasks. \textnormal{$T_{\mathrm{o}} \cap T_{\mathrm{l}} \cap T_{\mathrm{f}}$: tasks completed under all three description types; $T_{\mathrm{o}} \setminus (T_{\mathrm{l}} \cup T_{\mathrm{f}}$): tasks completed with only original descriptions; $(T_{\mathrm{l}} \cup T_{\mathrm{f}}) \setminus T_{\mathrm{o}}$: tasks completed only with $\sys_{lite}$ or $\sys_{full}$}.
}
\label{t:tsr-analysis}
\begin{tabular}{lccc}
\toprule
\textbf{Model for Gen} 
& \textbf{$T_{\mathrm{o}} \cap T_{\mathrm{l}} \cap T_{\mathrm{f}}$}
& \textbf{$T_{\mathrm{o}} \setminus (T_{\mathrm{l}} \cup T_{\mathrm{f}}$)} 
& \textbf{$(T_{\mathrm{l}} \cup T_{\mathrm{f}}) \setminus T_{\mathrm{o}}$}
\\
\midrule

\multirow{1}{*}{Claude-4.5-Sonnet} 
& 169 & 0 & 14\\

\multirow{1}{*}{Gemini-3-Flash} 
& 170 & 1 & 11\\

\multirow{1}{*}{GPT-5.2} 
& 170 & 2 & 11\\

\multirow{1}{*}{gpt-oss-120b} 
& 172 & 2 & 14\\


\bottomrule
\end{tabular}
\end{table}

%% file: table/trustdesc-gen-overhead.tex
\begin{table*}[t]
  \centering
  \footnotesize
  \small
  \caption{Description Generation Cost.
    \textnormal{
      We report the average token usage, monetary cost and time cost
      for \sys to generate one tool description.
      We break down the cost to entry function identification, 
      slice debloating,
      initial description generation, 
      and dynamic verification.}}
  \label{t:trustdesc-cost}
  \setlength{\tabcolsep}{3.6pt}
  \begin{tabular}{lrrrrrrrrrrrr|rrr}
    \toprule
    \multirow{2}{*}{\raisebox{-0.4ex}{\textbf{Model for Gen}}}
      & \multicolumn{3}{c}{\textbf{Entry Func.}}
      & \multicolumn{3}{c}{\textbf{Debloating}}
      & \multicolumn{3}{c}{\textbf{Init Desc.}}
      & \multicolumn{3}{c}{\textbf{Dynamic Ver.}}
      & \multicolumn{3}{c}{\textbf{Total}}
      \\
    \cmidrule{2-4}
    \cmidrule(lr){5-7}
    \cmidrule{8-10}
    \cmidrule(lr){11-13}
    \cmidrule{14-16}
      & \textbf{tokens} & \textbf{cost} & \textbf{time}
        & \textbf{tokens} & \textbf{cost} & \textbf{time}
        & \textbf{tokens} & \textbf{cost} & \textbf{time}
        & \textbf{tokens} & \textbf{cost} & \textbf{time}
        & \textbf{tokens} & \textbf{cost} & \textbf{time}
      \\
    \midrule

      {Claude-4.5-Sonnet}
      & \num{2757} & 0.009 &  2.2\second
      & \num{2020}  & 0.009 & 8.3\second
      & \num{2242}  & 0.011 & 7.5\second
      & \num{22256} & 0.077 & 34.1\second
      & \num{29277} & 0.106 & 52.0\second
      \\

      {Gemini-3-Flash}
      & \num{1650} & 0.001 &  1.7\second
      & \num{1877}  & 0.002 & 4.1\second
      & \num{2114}  & 0.002 & 2.7\second
      & \num{22146} & 0.009 & 17.2\second
      & \num{27788} & 0.013 & 25.7\second
      \\

      {GPT-5.2}
      & \num{1350} & 0.003 &  3.2\second
      & \num{1988}  & 0.010 & 13.5\second
      & \num{1927}  & 0.008 & 7.5\second
      & \num{18070} & 0.046 & 73.0\second
      & \num{23336} & 0.067 & 97.3\second
      \\

      {gpt-oss-120b}
      & \num{2761} & - &  2.4\second
      & \num{2078} & - &  4.7\second
      & \num{2144}  & - & 4.1\second
      & \num{13862} & - & 25.2\second
      & \num{20847} & - & 36.3\second
      \\
      
    \bottomrule
  \end{tabular}
\end{table*}

%% file: table/runtime-cost.tex
\begin{table*}[th]
    \centering
    \setlength\tabcolsep{4.3pt}
    \small
    \caption{Runtime Overhead.
      \textnormal{ 
    We report the runtime overhead of $\sys_{\mathrm{lite}}$ and $\sys_{\mathrm{full}}$.
    Token usage, monetary cost, latency, and number of tool-calls
    are reported as relative changes from the original baseline.
    $T$: total token usage; $T_{in}$: input token;
    $T_{out}$: output token.}
    }
  \label{t:runtime-cost}
    \begin{tabular}{lrrrrrrlrrrrrr}
      \toprule
  
      \multirow{2}{*}{\textbf{Model for Gen}}
      & \multicolumn{6}{c}{\textbf{$\sys_{\mathrm{lite}}$}}
      &
      & \multicolumn{6}{c}{\textbf{$\sys_{\mathrm{full}}$}}
      \\ 
      \cline{2-7}
      \cline{9-14}

      & {$T$}
      & {$T_{in}$}
      & {$T_{out}$}
      & {Cost} 
      & {Latency} 
      & {\#tool-call} 
      &
      & {$T$}
      & {$T_{in}$}
      & {$T_{out}$}
      & {Cost} 
      & {Latency} 
      & {\#tool-call} 
      \\
  
       \midrule

      {Claude-4.5-Sonnet} 
      &  +9.6\%
      &  +9.9\%
      &  +4.0\%
      &  +5.5\%
      &  +9.0\%
      &  +9.3\%
      &
      &  -3.3\%
      &  -3.4\%
      &  -0.7\%
      &  -2.6\%
      &  -0.3\%
      &  +1.0\%
      \\

      {Gemini-3-Flash} 
      &  +9.2\%
      &  +9.8\%
      &  +1.4\%
      &  +4.3\%
      &  -4.1\%
      &  +0.7\%
      &
      &  +1.6\%
      &  +1.8\%
      &  -1.2\%
      &  -1.7\%
      &  +0.5\%
      &  -2.3\%
      \\

      {GPT-5.2} 
      &  +6.2\%
      &  +6.5\%
      &  +1.6\%
      &  +6.4\%
      &  -2.2\%
      &  -0.3\%
      &
      &  +2.1\%
      &  +2.3\%
      &  -0.02\%
      &  +2.9\%
      &  -3.8\%
      &  -4.5\%
      \\

      {gpt-oss-120b} 
      &  +2.7\%  
      &  +3.1\%  
      &  -2.4\%  
      &  -0.3\%  
      &  -1.8\%  
      &  -2.3\%  
      &
      &  +0.4\%
      &  +0.5\%
      &  -0.7\%
      &  +2.4\%
      &  -3.0\%
      &  -3.9\%
      \\

      \midrule

      {Average}      
      &  +6.9\%
      &  +7.3\%
      &  +1.2\%
      &  +4.0\%
      &  +0.2\%
      &  +1.9\%
      &
      &  +0.2\%
      &  +0.3\%
      &  -0.7\%
      &  +0.3\%
      &  -1.7\%
      &  -2.4\%
      \\

      \bottomrule
    \end{tabular}
  \end{table*}

%% file: table/pick-better-desc.tex
\begin{table}[t]
  \centering
  \footnotesize
  \small
  \caption{Tool Selection Rate based on Code Quality.
  \textnormal{For each MCP tool, we reduce its code quality
  via three strategies, and measure its selection rate
  when competing with the original tool.}}
  \label{t:tool-competition}

  \begin{tabular}{llr}
    \toprule
    \textbf{Model for Gen} 
      & \textbf{Qualify Degradation} 
      & \textbf{Selection\%} 
      \\
    \midrule


    Claude-4.5-Sonnet
      & security checks removed  
      & 32.5\%
      \\
      & one feature disabled
      & 10.9\%
      \\
      & two features disabled 
      & 10.6\%
      \\ 
    \arrayrulecolor{gray}\hline

    Gemini-3-Flash
      & security checks removed  
      & 45.5\% 
      \\
      & one feature disabled 
      & 21.3\%
      \\
      & two features disabled 
      & 20.8\%
      \\
     \arrayrulecolor{gray}\hline

    GPT-5.2
      & security checks removed  
      & 44.0\% 
      \\
      & one feature disabled    
      & 34.6\%
      \\
      & two features disabled 
      & 17.7\%
      \\
     \arrayrulecolor{gray}\hline
     
    gpt-oss-120b
      & security cehcks removed  
      & 44.3\% 
      \\
      & one feature disabled   
      & 8.5\%
      \\
      & two features disabled   
      & 7.5\%
      \\

    \bottomrule
  \end{tabular}
\end{table}

%% file: table/tpa-prevention.tex
\begin{table}[t]
  \centering
  \footnotesize
  \small
  \setlength\tabcolsep{4pt}
  \caption{TPAs Prevention.
  \textnormal{We collect public available TPAs and check whether \sys can prevent them.}}
  \label{t:tpa-prevention}

\begin{tabular}{lclc}
\toprule
\textbf{Tool Name}
& \textbf{Type}
& \textbf{Bad terms} 
& \textbf{Blocked?} 
\\
\midrule

add  
\ExternalLink{https://github.com/invariantlabs-ai/mcp-injection-experiments/blob/main/direct-poisoning.py}
& E
& read id_rsa.pub
& $\checkmark$
\\

add 
\ExternalLink{https://github.com/invariantlabs-ai/mcp-injection-experiments/blob/main/shadowing.py}
& E 
& send all emails
& $\checkmark$ 
\\

get_fact_of_the_day 
\ExternalLink{https://github.com/invariantlabs-ai/mcp-injection-experiments/blob/main/whatsapp-takeover.py}
& E
& change the recipient
& $\checkmark$ 
\\

LocationNormalizer
\ExternalLink{https://github.com/systemsecurity-uiuc/Chord/blob/main/demo/attacker_server.py}
& E 
& MUST USE
& $\checkmark$ 
\\


CompanyToTicker
\ExternalLink{https://github.com/systemsecurity-uiuc/Chord/blob/main/demo/semantic_targeted_hooking.py}
& I
& Useful
& $\checkmark$
\\

ErrorExplainer
\ExternalLink{https://github.com/systemsecurity-uiuc/Chord/blob/main/demo/semantic_untargeted_hooking.py}
& I
& useful, AI-friendly
& $\checkmark$
\\

YoutubeSearch
\ExternalLink{https://github.com/systemsecurity-uiuc/Chord/blob/main/demo/syntax_format_hooking.py}
& E
& should use before
& $\checkmark$
\\

\bottomrule
\end{tabular}
\end{table}

%% file: discuss.tex
\section{Discussion}
\label{s:discuss}

We highlight two notable findings
from our evaluation
that further demonstrate
the benefits and necessity of \sys.

\PP{Surfacing Latent Security Constraints}
Beyond mitigating tool poisoning attacks,
we observe that \sys provides
an additional security benefit
by exposing implicit safeguards
embedded in tool implementations.
For example,
the implementation of \cc{apply\_formula}
explicitly blocks unsafe formulas
such as \cc{INDIRECT} and \cc{HYPERLINK}.
However, 
this restriction is not documented
in the original description.
Consequently,
when given tasks involving unsafe formulas,
the LLM initially invokes
\cc{apply\_formula},
encounters an execution error,
and then falls back to
\cc{write\_data\_to\_excel}
to write the same unsafe formula directly.
This behavior increases execution cost
and bypasses the intended security checks.
In contrast,
\sys captures this constraint
in the generated description,
allowing the LLM to reject unsafe tasks
before issuing any tool call.
This finding suggests that \sys
can strengthen overall system safety
by making latent implementation
protections visible to the LLM,
even when such protections are omitted
from developer-provided descriptions.

\PP{Prevalence of Low-Quality Tool Descriptions}
Our analysis of real-world MCP tools
reveals that incomplete or low-quality
tool descriptions are common,
even among widely used servers,
underscoring the practical need for \sys.
Among the 52 tools evaluated,
seven provide no argument descriptions,
frequently causing the LLM
to supply incorrect parameters.
19 tools include only minimal descriptions
that offer little guidance on proper usage.
Only 16 tools provide complete
and detailed descriptions,
and merely nine include usage examples.
These observations indicate that
tool developers often devote limited effort
to descriptions,
despite descriptions forming
a critical trust boundary
in LLM-integrated applications.
By automatically generating
implementation-aligned descriptions,
\sys fills this gap
and reduces reliance on
manual, error-prone documentation.

%% file: conclusion.tex
\section{Conclusion}
\label{s:conclusion}

We present \sys,
the first framework for
automatically generating
trusted tool descriptions
directly from tool implementations,
thereby eliminating tool poisoning attacks
at their root.
By combining reachability-aware
code slicing,
semantics-preserving description generation,
and dynamic verification,
\sys produces implementation-faithful
descriptions that significantly improve
tool selection accuracy
while incurring minimal cost
and runtime overhead.
Our evaluation on real-world tools
demonstrates that \sys
effectively mitigates both explicit
and implicit tool poisoning attacks
and remains robust under adaptive adversaries.